\documentclass[twocolumn,footinbib,floatfix,aps,pra,superscriptaddress]{revtex4-1}
\usepackage[usenames,dvipsnames]{xcolor} 
\usepackage[normalem]{ulem}

\usepackage{amsmath,amssymb}
\usepackage{amsfonts}
\usepackage{bm}
\usepackage{bbm}
\usepackage{color}
\usepackage{latexsym}
\usepackage[english]{babel}
\usepackage{times}
\usepackage{stmaryrd}
\usepackage{psfrag,graphicx}
\usepackage{subfigure}
\usepackage{natbib}
\usepackage{appendix}
\usepackage{enumitem}
\usepackage{bbold}

\definecolor{mygrey}{gray}{0.35}
\definecolor{myblue}{rgb}{0.2,0.2,0.8}
\definecolor{myzard}{cmyk}{0,0,0.05,0}
\definecolor{mywhite}{rgb}{1,1,1}
\definecolor{myred}{rgb}{0.9,0.1,0.}
\usepackage[colorlinks=true,citecolor=myblue,linkcolor=myred,urlcolor=myblue]{hyperref}

\newcommand{\e}{\operatorname{e}}
\renewcommand{\i}{\mathrm{i}}
\newcommand{\sinc}{\operatorname{sinc}}
\newcommand{\be}{\begin{equation}}
\newcommand{\ee}{\end{equation}}
\newcommand{\ben}{\begin{eqnarray}}
\newcommand{\een}{\end{eqnarray}}
\newcommand{\bes}{\begin{subequations}}
\newcommand{\ees}{\end{subequations}}
\newcommand{\bF}{\begin{figure}}
\newcommand{\eF}{\end{figure}}

\newcommand{\ket}[1]{\left| #1 \right>} 
\let\baraccent=\= 

\begin{document}
\title{Quantum enhanced estimation of a multi-dimensional field}

\author{Tillmann Baumgratz}
\affiliation{Clarendon Laboratory, Department of Physics, University of Oxford, OX1 3PU Oxford, United Kingdom}

\author{Animesh Datta}
\affiliation{Department of Physics, University of Warwick, CV4 7AL Coventry, United Kingdom}


\begin{abstract}
We present a  framework for the quantum enhanced estimation of multiple parameters corresponding to non-commuting unitary generators. Our formalism provides a recipe for the simultaneous estimation of all three components of a magnetic field. We propose a probe state that surpasses the precision of estimating the three components individually and discuss measurements that come close to attaining the quantum limit. Our study also reveals that too much quantum entanglement may be detrimental to attaining the Heisenberg scaling in quantum metrology.
\end{abstract}

\maketitle

\textit{Introduction:} As the elementary theory of nature, quantum mechanics sets the fundamental limit to precision of parameter estimation. On the flip side, quantum resources enable the estimation of parameters with a  precision surpassing that set by classical physics. This is the basis of the field of quantum enhanced sensing and metrology, and has been studied in great depth both theoretically and experimentally~\cite{Giovanetti2011,Toth2014,Demkowicz-Dobrzanski2014,Giovannetti2006}. Although most of these investigations have largely focussed on the estimation of a single phase parameter, more recently attention has been cast on the quantum-enhanced estimation of multiple parameters simultaneously~\cite{Genoni2013,humphreys2013,vidrighin2014,crowley2014,Yue2014,Zhang2014,Gao2014,Tsang2014,Kok2015}, and some early experiments have already been performed~\cite{Spagnolo2012}.

The motivations for studying quantum-enhanced multi-parameter estimation are manifold: First, while single phase estimation captures a wide range of scenarios~\cite{Giovannetti2004}, high-level applications such as microscopy, optical, electromagnetic, or gravitational field imaging, and spectroscopy intrinsically involve multiple parameters that should be estimated simultaneously. Secondly, while the quantum-enhanced limit for individual phase estimation can always be attained~\cite{Helstrom1976,Paris2009}, the measurements required to attain the quantum-enhanced limit for multiple parameters need not necessarily commute. This makes multi-parameter quantum-enhanced sensing a very interesting scenario for studying the limits of quantum measurements \cite{vidrighin2014,crowley2014}. Finally, multi-parameter quantum-enhanced sensing provides a novel paradigm for investigating the information processing capabilities of multi-partite or multi-mode quantum correlated states and measurements.

In this work, we study the problem of estimating a multi-dimensional field using a fixed number of particles. 
We first show that for a uniform field, the quantum enhancement to the precision of estimation is provided entirely by the two-particle reduced density matrix of the system, and that the attainability of the quantum enhancement is solely determined by the one-body reductions of the probe state.
We apply our methods to the simultaneous estimation of all the components of a magnetic field in three dimensions and show that this can be about three times better than estimating the components individually~\cite{Pang2014,Leibfried2004,Jones2009,Meyer2001}. Finally, we present a multi-partite quantum state achieving this advantage, and show how realistic measurements perform in attaining the multi-parameter quantum limit using matrix product state techniques~\cite{Fannes1992,Perez2007,Schollwoeck2011}.

\textit{Framework:} We consider the estimation of parameters governed by the Hamiltonian
\begin{equation}
\hat{H}(\boldsymbol{\varphi})=  \sum_{k=1}^{d} \varphi_k \hat{H}_{k}.
\label{eqn:Hfull}
\end{equation}
The parameters $\varphi_k \in \mathbb{R}$, $k=1,\ldots,d$, to be estimated  are the coefficients of a set of (not necessarily commuting) generators $\hat{H}_{k}.$ We assume that the $\hat{H}_k$ themselves do not depend on $\boldsymbol{\varphi}$. In addition to estimating a field in multiple dimensions simultaneously in free space, materials or biological samples,  this problem is equivalent to quantum-enhanced Hamiltonian tomography as it allows us to estimate unknown coefficients of the Hamiltonian in a suitable operator decomposition~\cite{Skotiniotis2015}. We note that earlier works have studied the estimation of parameters corresponding to unitary channels from information geometry~\cite{Fujiwara2001,Imai2007,Matsumoto2002} and representation theory~\cite{Bagan2001,Chiribella2005} perspectives, and have shown a Heisenberg scaling in their estimation.

A pure $N$-particle probe state $|\psi\rangle$ acquires the parameters via the unitary transformation $\hat{U}(\boldsymbol{\varphi}) = \e^{-\i\hat{H}(\boldsymbol{\varphi})}$ and we seek the best quantum strategy for the estimation of the parameters from the evolved probe state $|\psi_{\boldsymbol{\varphi}}\rangle = \hat{U}(\boldsymbol{\varphi})|\psi\rangle$. The performance of an estimator of $\boldsymbol{\varphi}$ is quantified in terms of the covariance matrix $\mathrm{Cov}[\boldsymbol{\varphi}]$. The quantum Cram{\'e}r-Rao bound~\cite{Helstrom1976,Paris2009} is a lower bound to the covariance matrix in terms of the quantum Fisher information matrix (QFIM), thus yielding an ultimate limit on the best possible precision of any (unbiased) estimator. For every specific set of positive operator valued measurements (POVM) $\{\hat{\Pi}_i\}$, one finds \cite{Paris2009}
\begin{equation}
M\mathrm{Cov}[\boldsymbol{\varphi}] \geq  \mathcal{F}(\boldsymbol{\varphi},\{\hat{\Pi}_{i}\})^{-1} \geq \mathcal{I}(\boldsymbol{\varphi})^{-1},
\label{eqn:QuantumCramerRaoBound}
\end{equation}
where the first inequality is the classical and the second inequality the quantum Cram{\'e}r-Rao bound, respectively. Here, $M$ is the number of times the overall experiment is repeated and $\mathcal{F}_{k,l}(\boldsymbol{\varphi},\{\hat{\Pi}_i\}) =  \sum_{n} \partial_{\varphi_k} p(n\vert \boldsymbol{\varphi}) \partial_{\varphi_l} p(n\vert \boldsymbol{\varphi})/p(n\vert \boldsymbol{\varphi})$, $k,l=1,\ldots,d$, denotes the Fisher information matrix (FIM) determined by the probabilities $p(n\vert \boldsymbol{\varphi}) = \langle\psi_{\boldsymbol{\varphi}}| \hat{\Pi}_{n} |\psi_{\boldsymbol{\varphi}}\rangle$. Further, $\mathcal{I}_{k,l}(\boldsymbol{\varphi}) =  \operatorname{Re}\bigl[\langle\psi _{\boldsymbol{\varphi}}| \hat{L}_{k}\hat{L}_{l}|\psi_{\boldsymbol{\varphi}}\rangle\bigr]$ is the QFIM, where, for pure probe states, the symmetric logarithmic derivative (SLD) $\hat{L}_{k}$ with respect to the parameter $\varphi_k$ is determined by
$\hat{L}_{k} = 2\left[ |\partial_{\varphi_k}\!\psi_{\boldsymbol{\varphi}}\rangle\langle \psi_{\boldsymbol{\varphi}}| + |\psi_{\boldsymbol{\varphi}}\rangle\langle\partial_{\varphi_{k}}\!\psi_{\boldsymbol{\varphi}}| \right]$  for all $k=1,\ldots,d$~\cite{Paris2009}.

While the classical Cram{\'e}r-Rao bound can always be saturated by, e.g., a maximum likelihood estimator~\cite{Braunstein1992}, the quantum limit (i.e., the second inequality in Eqn.~\eqref{eqn:QuantumCramerRaoBound}) may not be attainable in general. In a single parameter setting, the optimal measurements saturating the quantum Cram{\'e}r-Rao bound are given by the projectors onto the eigenvectors of the SLD. In the multi-parameter setting, however, the SLDs may not commute in general, leading to tradeoffs for the precisions of the individual estimators \cite{vidrighin2014,crowley2014}.

\textit{Formalism}:  For unitary time evolutions under Hamiltonians of the form of Eqn.~\eqref{eqn:Hfull} the QFIM can be expressed as the correlation matrix of the Hermitian operators defined by~\cite{Wilcox1967} (see Appendix~\ref{app:UnitaryMultiParameterEstimationQFIM})
\begin{equation}
\label{eqn:AOperatorGeneral}
\hat{A}_k(\boldsymbol{\varphi}) = \int_0^1\!\! d\alpha \, \e^{\i\alpha \hat{H}(\boldsymbol{\varphi})}\hat{H}_k\e^{-\i\alpha\hat{H}(\boldsymbol{\varphi})},
\end{equation}
leading to (suppressing the parameter $\boldsymbol{\varphi}$ in the arguments henceforth)
\begin{equation}
\mathcal{I}_{k,l} = 4\operatorname{Re}\!\big[ \langle \psi|  \hat{A}_k\hat{A}_l|\psi \rangle -  \langle\psi|\hat{A}  _k|\psi\rangle  \langle\psi|\hat{A}_l|\psi\rangle \big].
\label{eqn:QFIwithAoperator}
\end{equation}

We now restrict to the situation where the $N$ particles evolve under the one-particle Hamiltonian $\hat{h}^{[n]} = \sum_{k=1}^{d} \varphi_k \hat{h}_{k}^{[n]}$ for $n=1,\ldots,N$ (where the $\hat{h}_{k}^{[n]}$ are bounded), leading to the global Hamiltonian 
\begin{equation}
\hat{H}(\boldsymbol{\varphi}) = \sum_{n=1}^{N} \hat{h}^{[n]} =  \sum_{k=1}^{d} \varphi_k \sum_{n=1}^{N} \hat{h}_{k}^{[n]} \equiv  \sum_{k=1}^{d} \varphi_k \hat{H}_{k}.
\label{eqn:Hsmall}
\end{equation}
With this, the operators defined in Eqn.~\eqref{eqn:AOperatorGeneral} simplify to $\hat{A}_k(\boldsymbol{\varphi})\equiv \sum_{n=1}^{N} \! \hat{a}_k^{[n]},$ where 
\begin{equation}
\label{eqn:AOperatorLocal}
\hat{a}_k^{[n]} = \int_0^1\!\! d\alpha \, \e^{\i\alpha \hat{h}^{[n]}(\boldsymbol{\varphi})}\hat{h}_k^{[n]}\e^{-\i\alpha\hat{h}^{[n]}(\boldsymbol{\varphi})}
\end{equation}
are Hermitian operators acting only on particle $n$. Incorporating this in Eqn.~\eqref{eqn:QFIwithAoperator} simplifies the QFIM to
\begin{widetext}
\begin{eqnarray}
\label{eqn:SimplificationOfQFI}
\mathcal{I}_{k,l} &=& 4\!\!\!\sum_{n,m=1}^{N}\!\operatorname{Re}\!\Big[\langle\psi|\hat{a}_{k}^{[n]}\hat{a}_{l}^{[m]}|\psi\rangle - \langle\psi|\hat{a}_{k}^{[n]}|\psi\rangle  \langle\psi|\hat{a}_{l}^{[m]}|\psi \rangle\Big] \\ \nonumber
&=& 4\sum_{n}\operatorname{Re}\!\!\Big[\operatorname{Tr}\!\big[\hat{\varrho}^{[n]}\hat{a}_{k}^{[n]}\hat{a}_{l}^{[n]}\big] \!-\! \operatorname{Tr}\!\big[\hat{\varrho}^{[n]}\hat{a}_{k}^{[n]}\big] \operatorname{Tr}\!\big[\hat{\varrho}^{[n]}\hat{a}_{l}^{[n]}\big]\Big] \!+\!4\! \sum_{n\neq m}\! \operatorname{Re}\!\!\Big[\operatorname{Tr}\!\big[\hat{\varrho}^{[n,m]}\hat{a}_{k}^{[n]} \otimes\hat{a}_{l}^{[m]}\big] \!-\! \operatorname{Tr}\!\big[\hat{\varrho}^{[n]}\hat{a}_{k}^{[n]}\big] \operatorname{Tr}\!\big[\hat{\varrho}^{[m]}\hat{a}_{l}^{[m]}\big]\Big],
\end{eqnarray}
\end{widetext}
where $\hat{\varrho}^{[n,m]} = \operatorname{Tr}_{\backslash\{n,m\}}[|\psi\rangle\langle\psi|]$ denotes the reduced density matrix to sub-systems $n,m$. The first sum has $N$ terms, and the second has $\mathcal{O}(N^2)$ terms. The latter points to the origin of the quadratic scaling in $N$ in quantum-enhanced metrology, and shows that quantum advantage in metrology can be no more than quadratic in the considered scenario of Eqn.~\eqref{eqn:Hsmall} if the variance of the estimator is the quantifier of precision. 

For estimating a uniform field, the phase parameters are identical across the system (although corresponding to non-commuting generators) and we can restrict ourselves to permutationally invariant quantum states, i.e., states that are invariant under any exchange of its constituents. The one- and two-particle reduced density matrices are then given by $\hat{\varrho}^{[n]}=\hat{\varrho}^{[1]}$ and $\hat{\varrho}^{[n,m]}=\hat{\varrho}^{[2]}$ for all $n,m$, respectively. Under the restriction of permutationally invariant states, the QFIM simplifies to
\begin{equation}
\mathcal{I} = 4N \mathcal{I}^{[1]} + 4N(N-1) \mathcal{I}^{[2]},
\label{eqn:QFIForPIPureStates}
\end{equation}
where 
\begin{equation}
\mathcal{I}^{[1]}_{k,l} = \operatorname{Re}\!\big[ \operatorname{Tr}\!\big[ \hat{\varrho}^{[1]} \hat{a}_{k}\hat{a}_{l}\big]\big] - \operatorname{Tr}\!\big[ \hat{\varrho}^{[1]}\hat{a}_{k}\big] \operatorname{Tr}\big[ \hat{\varrho}^{[1]}\hat{a}_{l}\big] 
\label{eqn:QFIForPIPureStatesPartI}
\end{equation}
only depends on the one-particle reduced density matrix and 
\begin{equation}
\mathcal{I}^{[2]}_{k,l} = \operatorname{Tr}\!\big[ \hat{\varrho}^{[2]} \hat{a}_{k}\otimes\hat{a}_{l}\big] - \operatorname{Tr}\!\big[ \hat{\varrho}^{[1]}\hat{a}_{k}\big] \operatorname{Tr}\!\big[ \hat{\varrho}^{[1]}\hat{a}_{l}\big]
\label{eqn:QFIForPIPureStatesPartII}
\end{equation}
depends on the two-particle reduced density matrix.

Eqn.~\eqref{eqn:QFIForPIPureStates} highlights several interesting physical aspects of quantum-enhanced metrology: 
First, note that $\mathcal{I}^{[1]}$ can be bounded independently of $\hat{\varrho}^{[1]}$. This immediately shows that the archetypal quadratic scaling of quantum-enhanced sensing arises solely from the two-particle reduced terms. For instance, let the probe state be $|\psi\rangle = |\phi\rangle^{\otimes N}$, i.e., permutationally invariant and separable. Then, $\hat{\varrho}^{[2]} = \hat{\varrho}^{[1]}\otimes \hat{\varrho}^{[1]}$ such that $\mathcal{I}^{[2]} = 0$ and the QFIM only scales linearly in $N$, i.e., $\mathcal{I} = N \mathcal{I}^{[1]}$. Thus, Eqn.~\eqref{eqn:QFIForPIPureStates} implies that in permutationally invariant systems quantum correlations are necessary for achieving a quadratic scaling in the number of probe states $N$---the so-called Heisenberg scaling. Note that the latter reasoning also applies to quantum states that are not permutationally invariant, as can be seen by Eqn.~\eqref{eqn:SimplificationOfQFI}. 
Further, for probe states of the form $|\psi\rangle = |\phi\rangle^{\otimes N}$, the QFIM satisfies $\operatorname{rank}[\mathcal{I}] \leq 2(D-1)$ where $D$ is the dimension of the local Hilbert space (e.g., $D=2$ for two-level systems, see Appendix~\ref{app:TheQFIForProductProbeStates} for details) such that if the number of parameters exceeds $2(D-1)$, i.e., $d>2(D-1)$, a simultaneous estimation of all parameters necessarily fails due to a lack of information for all parameters in the QFIM.
Finally, if both the one- and two-particle reduced states are maximally mixed, the Heisenberg scaling is lost. To see this, note that $\hat{\varrho}^{[1]} = \mathbbm{1}_2/2$ (where $\mathbbm{1}_k$ is the $k \times k$ identity matrix) implies
$\mathcal{I}^{[2]} = \operatorname{Tr}\big[ \hat{\varrho}^{[2]} \hat{a}_{k}\otimes\hat{a}_{l}\big] - \operatorname{Tr}\big[ \hat{a}_{l}\big] \operatorname{Tr}\big[ \hat{a}_{l}\big] /4,$ which vanishes if $\hat{\varrho}^{[2]} = \mathbbm{1}_4/4$. This is an example where too much entanglement harms the quantum advantage of exploiting $N$ particles in parallel.

\textit{Attaining the quantum limit:}  Saturating the quantum Cram\'er-Rao bound and attaining the QFIM is the next important part of quantum-enhanced sensing. This is particularly interesting for multi-parameter estimation since the SLDs corresponding to the different parameters need not commute.  We show in Appendix~\ref{app:UnitaryMultiParameterEstimationSLDs} that for a purely unitary evolution, the QFIM is saturated if the expectation value of the commutator of the SLDs vanishes for all pairs~\cite{Matsumoto2002}, i.e., 
\begin{equation}
\langle\psi_{\boldsymbol{\varphi}}|\hat{L}_{k}\hat{L}_{l} - \hat{L}_{l}\hat{L}_{k}|\psi_{\boldsymbol{\varphi}}\rangle \equiv 8\i\operatorname{Im}\!\!\big[ \langle\psi|\hat{A}_{k}\hat{A}_{l}|\psi\rangle \big] = 0.
\label{eqn:ExpectationValueOfCommutatorSLDs}
\end{equation}
For permutation invariant systems, this reduces to $8\i N \operatorname{Im}\!\!\left[ \operatorname{Tr}\big[ \hat{\varrho}^{[1]} \hat{a}_{k}\hat{a}_{l} \big] \right] = 0$ for all $k,l$. 
It is interesting to note that while the quantum-enhanced scaling is governed entirely by the two-particle reduced density matrices (see Eqn.~(\ref{eqn:SimplificationOfQFI})), the attainability of this bound is determined solely by the one-particle term (for a general proof, see Appendix~\ref{app:UnitaryMultiParameterEstimationSLDs}). The expectation value vanishes, for instance, for permutationally invariant pure probe states $|\psi\rangle$ with $\hat{\varrho}^{[1]} = \mathbbm{1}_2/2$~\cite{remark1}. This is a sufficient but not necessary condition for the expectation of the commutator to vanish and gives a rather simple mathematical condition for the quantum Cram{\'e}r-Rao bound to be saturated. It is an instance of the \textit{local suppression} of the non-commutativity of the generators using quantum correlations~\cite{Fujiwara2001}.

More generally, when the expectation values of all commutators of the SLDs vanish and the QFIM is of full rank, the eigenvectors of the $d$ distinct SLDs lie in a subspace of dimension $d+1$ allowing for the construction of a POVM that saturates the quantum Cram{\'e}r-Rao bound. We prove this assertion in Appendix~\ref{app:UnitaryMultiParameterEstimationSLDs} and, further, provide a procedure for constructing such a POVM that saturates the quantum Cram{\'e}r-Rao bound. Note that for commuting generators $\langle\psi|\hat{A}_{k}\hat{A}_{l}|\psi\rangle \in\mathbb{R},$ such that the quantum Cram{\'e}r-Rao bound can always be saturated given the QFIM is not rank deficient (see also~\cite{Matsumoto2002}).

\textit{Estimating a magnetic field in three dimensions}: 
We now apply our formalism to the task of estimating the components of a magnetic field in three dimensions simultaneously using two-level systems. Potential systems could include trapped ions, NV centres, or doped spins in semiconductors~\cite{Richerme2014,Jurcevic2014,Warner2013,Albrecht2014,Lange2011}. The Hamilton operator for this system is given by (with $d=3$, see Appendix~\ref{app:dg3} for a discussion of $d>3$)
\begin{equation}
\hat{h} = \boldsymbol{\hat{\mu}}\cdot \boldsymbol{B} = \sum_{k=1}^{3} \hat{\mu}_k B_{k} = \sum_{k=1}^{3} \frac{\mu}{2}B_{k}\hat{\sigma}_{k} := \sum_{k=1}^{3}\varphi_k \hat{\sigma}_{k},
\label{eqn:OneParticleHamiltonianFor3BField}
\end{equation}
where the magnetic moment $\hat{\mu}_k = \mu \,\hat{\sigma}_k / 2$ is proportional to the spin, $\{\hat{\sigma}_{k}\}$ denotes the unnormalised Pauli operators, and $\varphi_k=\mu B_k/2$. To develop the intuition for estimating the magnetic field in three dimensions simultaneously, we start with the estimation of a magnetic field pointing solely along one of the specific directions X, Y, or Z. It is well known that a GHZ-type state (see Appendix~\ref{app:SingleParameterEstimation})
\begin{equation}
\ket{\Phi_k} = \left( |\phi^{+}_{k}\rangle^{\otimes N} + |\phi^{-}_{k}\rangle^{\otimes N} \right)/ \sqrt{2}
\end{equation}
achieves the quantum Cram{\'e}r-Rao bound, where $|\phi^{\pm}_{k}\rangle$ is the eigenvector of the Pauli operator $\hat{\sigma}_k$ corresponding to the eigenvalue $\pm1$ ($k=1,2,3$ corresponding to the X, Y, and Z direction). These states are permutationally invariant with one- and two-particle reduced density matrices $\hat{\varrho}^{[1]}_{k} = \mathbbm{1}_2/2$ and $\hat{\varrho}^{[2]}_{k} = ( |\phi^{+}_{k},\phi^{+}_{k}\rangle\langle \phi^{+}_{k},\phi^{+}_{k}| \!+\! |\phi^{-}_{k},\phi^{-}_{k}\rangle\langle \phi^{-}_{k},\phi^{-}_{k}| )/2 = ( \mathbbm{1}_2\otimes\mathbbm{1}_2 + \hat{\sigma}_{k}\otimes \hat{\sigma}_{k})/4$, respectively. Now, for the simultaneous estimation of all three components, an obvious candidate is 
\begin{equation}
|\psi\rangle = \mathcal{N}\!\left( \e^{\i\delta_1}|\Phi_1\rangle + \e^{\i\delta_2}|\Phi_2\rangle + \e^{\i\delta_3}|\Phi_3\rangle \!\right),
\label{eqn:SpecificProbeState}
\end{equation}
where $\mathcal{N}$ is the normalisation constant and $\{\delta_k\}$ adjustable local phases. Now, for even $N$ and appropriate $\delta_k$ the quantum Cram{\'e}r-Rao bound is achievable with this state since $\hat{\varrho}^{[1]} = \mathbbm{1}_2/2$~ \cite{remark2}. It remains to show that it is also capable of achieving the Heisenberg scaling. To simplify our calculations, we henceforth restrict ourselves to systems that consist of $N\equiv 8n$, $n\in\mathbb{N}$, particles (and $\delta_k=0$ for all $k$) but note that this is no limitation of our model as indicated by the numerical simulations presented below. For $N\equiv 8n$, $n\in\mathbb{N}$, the two-particle reduced density matrix of $|\psi\rangle$ is an equal mixture of the GHZ-type states in all directions and given by
\begin{equation}
\hat{\varrho}^{[2]} = \frac{1}{3} \sum_{k=1}^{3}\hat{\varrho}^{[2]}_k = \frac{1}{4} \mathbbm{1}_2\otimes\mathbbm{1}_2 + \frac{1}{12} \sum_{k=1}^{3} \hat{\sigma}_{k}\otimes\hat{\sigma}_{k}.
\label{eqn:TwoBodyReductionG} 
\end{equation}
For any other $N,$ we show in Appendix~\ref{app:Reduceddensitymatricesoftheprobestate} that the difference from the form of $\hat{\varrho}^{[2]}$ in Eqn.~\eqref{eqn:TwoBodyReductionG} is exponentially small in $N$. For a probe state with marginals $\hat{\varrho}^{[1]}=\mathbbm{1}_2/2$ and $\hat{\varrho}^{[2]}$ given above, the QFIM is (see Appendix~\ref{app:DerivationQFI} and Ref.~\cite{Imai2007} which shows the same scaling)
\begin{equation}
\mathcal{I}_{k,l} \!=\! \frac{4}{3}N(N+2)\!\left[ (1\!-\!\sinc^2[\xi]) \eta_k\eta_l + \delta_{k,l} \sinc^2[\xi]\right],
\label{eqn:QuantumFisherInformationMatrix}
\end{equation}
where $\sinc[\xi] = \sin[\xi]/\xi$ with 
$\xi = \sqrt{\varphi_1^2 + \varphi_2^2 + \varphi_3^2}$  and $\eta_k = \varphi_k/\sqrt{\varphi_1^2 + \varphi_2^2 + \varphi_3^2}$
for all $k$. Note that, in the limit of $\varphi_k\rightarrow 0$ for $k=1,2,3$, the QFIM is diagonal, i.e., $\mathcal{I}_{k,l} = \frac{4}{3}N(N+2) \delta_{k,l}$.
As our quantum Cram{\'e}r-Rao bound can be saturated, the minimal total variance for estimating the three {components of the magnetic field simultaneously is given by
\begin{equation}
|\Delta \boldsymbol{\varphi}^{\text{sim}}_{\text{ent}}|^2 = \sum_{k=1}^{3} \Delta\varphi_{k}^2 = \operatorname{Tr}[\operatorname{Cov}(\boldsymbol{\varphi})] = \operatorname{Tr}[\mathcal{I}^{-1}(\boldsymbol{\varphi})].
\end{equation}
Since the QFIM in Eqn.~\eqref{eqn:QuantumFisherInformationMatrix} is the sum of a rank one matrix and a rescaled identity, its eigenvalues can be read off directly as $\lambda_{1} = 4N(N+2)/3$ and $ \lambda_{2,3} = 4N(N+2)\sinc^2[\xi]/3$ leading to 
\begin{equation}
|\Delta \boldsymbol{\varphi}^{\text{sim}}_{\text{ent}}|^2= \frac{3+6/\sinc^2[\xi]}{4N(N+2)}.
\label{eqn:TotalSimultaneousEstimationVariance}
\end{equation}

Let us now compare three different scenarios for the estimation of $\boldsymbol{\varphi}$: (i) A classical strategy of using only pure product states, (ii) a quantum strategy where the parameters are estimated individually (as depicted in Fig.~\ref{fig:NumericalResults}~(a)), and (iii) the simultaneous estimation of the parameters, as shown in Fig.~\ref{fig:NumericalResults}~(b). To obtain a fair comparison amongst (i)-(iii), we use exactly $N$ particles to estimate all three cases. As for all scenarios the QFI depends on the values of $\varphi_{k}$, we restrict ourselves to a setting with $\varphi_{k}\rightarrow 0$, $k=1,\ldots,3$.
For scenario (i), the strategy is to divide the set of $N$ particles into three blocks of length $n=N/3$ and, on the $k^{\text{th}}$ block, to prepare a product state that allows for the estimation of $\varphi_k$. This is due to the impossibility of estimating 3 parameters simultaneously using a pure and permutationally invariant product state, as shown by the singularity of the QFIM (Appendix~\ref{app:TheQFIForProductProbeStates} shows that its rank is 2). 
The maximal QFI for each block (see Appendix~\ref{app:SingleParameterEstimation}) is equal to 
$\mathcal{I}_{k} = n(\lambda_{\text{max}}(\hat{a}_k)-\lambda_{\text{min}}(\hat{a}_k))^2$ where $\lambda_{\text{max/min}}(\hat{a}_{k})$ denotes the maximal/minimal eigenvalue of $\hat{a}_{k}$ such that
\begin{equation*}
(\lambda_{\text{max}}(\hat{a}_k)-\lambda_{\text{min}}(\hat{a}_k))^2 = 4\left[ (1- \sinc^2[\xi])\eta_k^2 + \sinc^2[\xi] \right]
\end{equation*}
with $(\lambda_{\text{max}}(\hat{a}_k)-\lambda_{\text{min}}(\hat{a}_k))^2 \rightarrow 4$ for $\varphi_k\rightarrow 0$, $k=1,2,3$. Further, $\Delta\varphi_k^2 = 1/\mathcal{I}_{k}$ and thus we find for the individual estimation of all parameters using separable states 
\begin{equation}
|\Delta\boldsymbol{\varphi}_{\text{sep}}^{\text{ind}}|^2 = \sum_{k=1}^{3} \Delta\varphi_{k}^2  \rightarrow \frac{9}{4N}. 
\end{equation}  

Secondly, for a quantum strategy exploiting entangled states where we estimate the parameters individually we, again, divide the chain of $N$ particles into three blocks. Next, on the $k^{\text{th}}$ block, one prepares a GHZ-type state in the $\hat{a}_{k}$ basis. Recall that for each block $\mathcal{I}_k = n^2(\lambda_{\text{max}}(\hat{a}_k)-\lambda_{\text{min}}(\hat{a}_k))^2$ (see Appendix~\ref{app:SingleParameterEstimation}) such that with $\Delta\varphi_k^2 = 1/\mathcal{I}_{k}$ one finds
\begin{equation}
|\Delta \boldsymbol{\varphi}^{\text{ind}}_{\text{ent}}|^2 =  \sum_{k=1}^{3} \Delta\varphi_{k}^2 \rightarrow \frac{27}{4N^2}. 
\end{equation}
Thirdly, for the simultaneous estimation of the parameters we have (see Eqn.~\eqref{eqn:TotalSimultaneousEstimationVariance})
\begin{equation}
|\Delta \boldsymbol{\varphi}^{\text{sim}}_{\text{ent}}|^2 \rightarrow \frac{9}{4N(N+2)}.
\end{equation}
Hence, it is possible to design quantum probes (those with a completely mixed one-particle reduced marginal and two-particle reduced marginal given by Eqn.~\eqref{eqn:TwoBodyReductionG}) for magnetic field estimation such that estimating the three components simultaneously is about three times better than estimating them individually.
Overall, $|\Delta \boldsymbol{\varphi}^{\text{sim}}_{\text{ent}}|^2 \leq |\Delta \boldsymbol{\varphi}^{\text{ind}}_{\text{ent}}|^2 \leq |\Delta \boldsymbol{\varphi}_{\text{sep}}^{\text{ind}}|^2$ for all $N \geq 3.$ This is illustrated in Fig.~\ref{fig:NumericalResults} (c) where the results are obtained numerically using matrix product state techniques~\cite{Fannes1992,Perez2007,Schollwoeck2011} to also account for system sizes $N\neq 8n$. Note that the 3-fold improvement is not proven to be optimal, although we expect it to be so.  This is a more general behaviour than the findings of \cite{humphreys2013}, where the generators were commuting while here they are not.

\textit{Classical Fisher information:} We have already shown (Appendix~\ref{app:UnitaryMultiParameterEstimationSLDs}) that there is a POVM that achieves the multi-parameter quantum Cram{\'e}r-Rao bound. The so-constructed POVM contains as one element the projector onto the time-evolved probe state, i.e., $\hat{U}(\boldsymbol{\varphi})|\psi\rangle$. While this set theoretically achieves the bound, it may not be very appealing from an experimental perspective.  Hence, let us finally discuss some realistic measurements. In particular, we consider two sets of POVMs: $\hat{\Pi}_k^{(1)}$, $k=1,\ldots,4$, contains the three projectors
\begin{equation*}
\hat{\Pi}_k^{(1)} \!=\! |\Psi_k\rangle\langle\Psi_k| \text{ with } |\Psi_{k}\rangle \!=\! \left(|\phi^{+}_{k}\rangle^{\otimes N} + \e^{\i\delta_k}\! |\phi^{-}_{k}\rangle^{\otimes N} \right)\!/\sqrt{2}
\end{equation*}
together with the element guaranteeing normalisation $\hat{\Pi}_{4}^{(1)} = \mathbbm{1} - \sum_{k=1}^{3}\hat{\Pi}_{k}^{(1)}$. 
Note that for even $N$ and appropriate $\delta_k$ these operators indeed form a valid set of POVMs \cite{remark3}. Further, $\hat{\Pi}_{k,\pm}^{(2)}$, $k=1,\ldots,3$, is determined solely by expectation values of simple Pauli strings, i.e., 
\begin{equation*}
\hat{\Pi}^{(2)}_{k,\pm} = \left( \mathbb{1} \pm \hat{\sigma}_{k}^{\otimes N} \right)/6.
\end{equation*}
Again, we use matrix product state techniques to compute the classical Fisher information for these POVMs. This enables us to analyse the scaling up to very large system sizes, as shown in Fig.~\ref{fig:NumericalResults} (c). Note that the precision for both these POVMs is best for  $N=8n$ and the considered probe state.

\begin{figure}[tb]
	\centering
	\includegraphics[width=1.0\columnwidth]{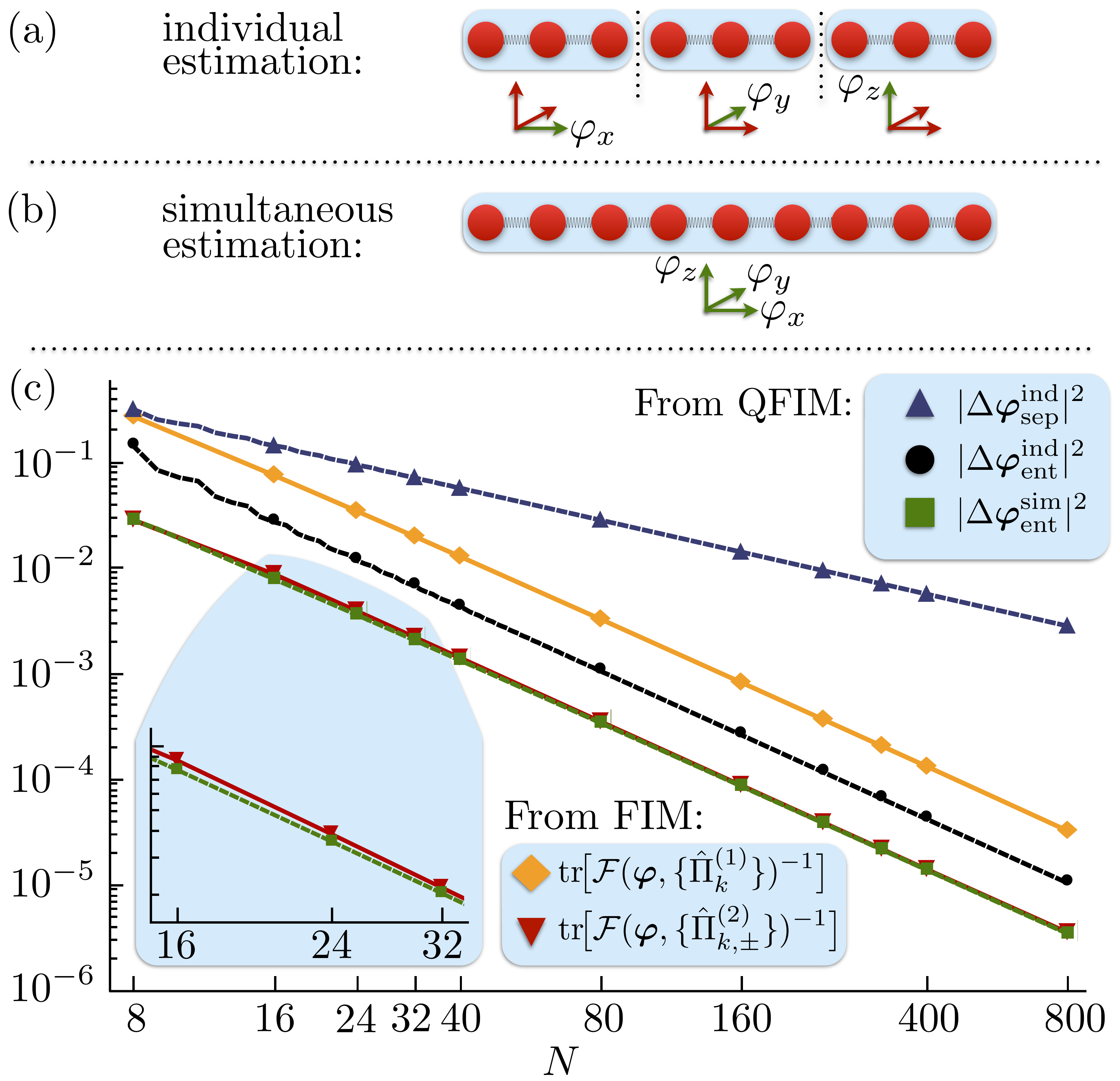}
	\caption{(a): Individual estimation of the phases can be achieved by dividing the chain into sub-blocks, each of which is used to estimate only a single parameter. This is contrasted by the simultaneous estimation of the parameters depicted in (b). (c): Log-Log plot for the estimation of the three directions of a magnetic field with parameters $\varphi_1=10^{-4}$, $\varphi_2=2\times 10^{-4}$, and $\varphi_3 = 3\times 10^{-4}$. We show the QFIM for the three different scenarios described in the main text, as well as the result obtained for the FIM for the two considered POVMs. Note that for the QFIM results we computed the total variance for all $N$, while for the FIM results only for the values of $N$ emphasised with a marker. }
\label{fig:NumericalResults}
\end{figure}

\textit{Conclusions:} We have obtained the quantum limits for the simultaneous estimation of parameters corresponding to non-commuting unitary generators. We applied our methods to the simultaneous estimation of all three components of a magnetic field in space. The results suggest that estimating the phases simultaneously improves the sensitivity by a factor of $d=3$, in consonance with earlier results with commuting generators~\cite{humphreys2013}. Future extensions of our results could include, amongst others, a combination of commuting and non-commuting generators, and the inclusion of decoherence. Another direction could be the search for optimal probe states and more tractable measurements for specific physical systems such as trapped ions or vacancy centres in diamond.

\textit{Acknowledgements:} This work was supported by the UK EPSRC (EP/K04057X/1, EP/M01326X/1, EP/M013243/1).

\appendix
\section{\label{app:UnitaryMultiParameterEstimationQFIM}Unitary multi-parameter estimation: \\ SLDs and QFIM}
In the first section of the appendix, we set out to find an expression for the symmetric logarithmic derivatives (SLDs) together with the quantum Fisher information matrix (QFIM) for the setting discussed in the main text. For this, we restrict ourselves to unitary channels  where the to-be-estimated parameters $\varphi_k \in \mathbb{R}$, $k=1,\ldots,d$, are the coefficients of a set of (not necessarily commuting) generators $\hat{H}_{k}$, i.e., we consider unitaries of the form
\begin{equation}
\hat{U}(\boldsymbol{\varphi}) = \e^{-\i \hat{H}(\boldsymbol{\varphi})} = \e^{-\i \sum_{k=1}^{d} \varphi_k \hat{H}_{k}},
\end{equation}
where $\hat{H}_{k}^{\dagger} = \hat{H}_{k}$ for all $k=1,\ldots,d$ and $\boldsymbol{\varphi}\in\mathbb{R}^{d}$ with $[\boldsymbol{\varphi}]_k = \varphi_k$. Further, note that the $\hat{H}_k$ do not depend on the parameters $\boldsymbol{\varphi}$. For a pure probe state $|\psi\rangle$ and purely unitary evolution, the SLDs are given by~\cite{Paris2009}
\begin{equation}
\hat{L}_{k} = 2\big[  |\partial_{\varphi_k} \psi_{\boldsymbol{\varphi}}\rangle \langle \psi_{\boldsymbol{\varphi}} | +  |\psi_{\boldsymbol{\varphi}}\rangle \langle \partial_{\varphi_k}\psi_{\boldsymbol{\varphi}} | \big],
\label{eqn:unitarySLDAppendix}
\end{equation}
where $|\partial_{\varphi_k}\psi_{\boldsymbol{\varphi}}\rangle = [\partial_{\varphi_k}\hat{U}(\boldsymbol{\varphi})]|\psi\rangle$ denotes the partial derivative of $|\psi_{\boldsymbol{\varphi}}\rangle$ with respect to the parameter $\varphi_k$. Now, recall that~\cite{Wilcox1967} (see \cite{Pang2014} for another application in quantum metrology)
\begin{equation}
\frac{\partial \e^{-\i \hat{H}(\boldsymbol{\varphi})}}{\partial \varphi_k} \!=\! -\i \!\int_0^1 \!\!\!\! d\alpha  \e^{-\i(1-\alpha)\hat{H}(\boldsymbol{\varphi})} \!\frac{\partial \hat{H}(\boldsymbol{\varphi})}{\partial \varphi_k} \e^{-\i \alpha \hat{H}(\boldsymbol{\varphi})}, 
\end{equation}
i.e., 
\begin{equation}
\frac{\partial}{\partial \varphi_k} |\psi_{\boldsymbol{\varphi}}\rangle = \frac{\partial}{\partial \varphi_k} \hat{U}(\boldsymbol{\varphi}) |\psi\rangle =   \hat{U}(\boldsymbol{\varphi}) \hat{O}_k(\boldsymbol{\varphi}) |\psi\rangle
\label{eqn:DerivativeWithAOperators}
\end{equation}
with the skew-Hermitian operator
\begin{equation}
\hat{O}_k(\boldsymbol{\varphi}) \!=\!-\i\hat{A}_k(\boldsymbol{\varphi})\!=\! -\i\int_0^1\!\!\! d\alpha \, \e^{\i\alpha \hat{H}(\boldsymbol{\varphi})}\hat{H}_k\e^{-\i\alpha\hat{H}(\boldsymbol{\varphi})}
\label{eqn:TheAOperators}
\end{equation}
and where we defined $\hat{A}_k(\boldsymbol{\varphi}) = \i\hat{O}_k(\boldsymbol{\varphi})$. With Eqns.~\eqref{eqn:unitarySLDAppendix} and \eqref{eqn:DerivativeWithAOperators} one finds 
\begin{equation}
\begin{split}
\hat{L}_k &= 2 \hat{U}\bigl[ \hat{O}_k|\psi\rangle\langle\psi| + |\psi\rangle\langle\psi|\hat{O}_k^{\dagger}\bigr] \hat{U}^{\dagger} \\
&= 2\i \hat{U}\bigl[ |\psi\rangle\langle\psi|,\hat{A}_{k}\bigr] \hat{U}^{\dagger},
\end{split}
\label{eqn:SLDForSetting}
\end{equation}
where $[\hat{X},\hat{Y}]$ denotes the commutator of the operators $\hat{X}$ and $\hat{Y}$, respectively. Next, let us consider the QFIM. For unitary time evolutions it is given by \cite{Paris2009,humphreys2013}
\begin{equation*}
\mathcal{I}_{k,l}(\boldsymbol{\varphi}) \!=\! 4 \operatorname{Re}\!\left[ \langle \partial_{\varphi_k}\!\psi_{\boldsymbol{\varphi}} |\partial_{\varphi_l}\!\psi_{\boldsymbol{\varphi}}\rangle  \!-\!   \langle \partial_{\varphi_k}\!\psi_{\boldsymbol{\varphi}} |\psi_{\boldsymbol{\varphi}}\rangle  \langle \psi_{\boldsymbol{\varphi}} |\partial_{\varphi_l}\!\psi_{\boldsymbol{\varphi}}\rangle \right]\!.
\end{equation*}
With this, Eqn.~\eqref{eqn:DerivativeWithAOperators} allows us to write the QFIM in terms of the correlation matrix of the operators $\{\hat{A}_k(\boldsymbol{\varphi})\}$. One finds
\begin{equation}
\mathcal{I}_{k,l}(\boldsymbol{\varphi}) = 4 \operatorname{Re}\!\big[ \langle \psi|  \hat{A}_k\hat{A}_l|\psi \rangle \!-\!  \langle\psi|\hat{A}_k|\psi\rangle  \langle\psi|\hat{A}_l|\psi\rangle \big].
\label{eqn:QFIwithAoperator_Appendix}
\end{equation}
Note that we omitted the explicit dependency of the operators  on the parameters $\boldsymbol{\varphi}$. Although the process is unitary the QFIM may depend on the parameters $\boldsymbol{\varphi}$, i.e., $\mathcal{I} = \mathcal{I}(\boldsymbol{\varphi})$.  Further, in general, we have $[\hat{A}_k(\boldsymbol{\varphi}),\hat{A}_l(\boldsymbol{\varphi})]\neq 0$.

\section{\label{app:TheQFIForProductProbeStates}The QFIM for product probe states}
In this section of the appendix, we prove an upper bound on the rank of the QFIM for separable probe states. Recall 
\begin{equation}
\mathcal{I} = 4N \mathcal{I}^{[1]} + 4N(N-1) \mathcal{I}^{[2]},
\end{equation}
where 
\begin{equation}
\mathcal{I}^{[2]}_{k,l} =\operatorname{Tr}\!\big[ \hat{\varrho}^{[2]} \hat{a}_{k}\otimes\hat{a}_{l}\big] - \operatorname{Tr}\!\big[ \hat{\varrho}^{[1]}\hat{a}_{l}\big] \operatorname{Tr}\!\big[ \hat{\varrho}^{[1]}\hat{a}_{l}\big] = 0 
\end{equation}
for product probe states, i.e., states of the form $|\psi\rangle = |\phi\rangle^{\otimes N}$ where $|\phi\rangle\in\mathbb{C}^{D}$. Note that for these states $\hat{\varrho}^{[1]} = |\phi\rangle\langle\phi|$ such that
\begin{eqnarray}
\mathcal{I}^{[1]}_{k,l} &=&  \operatorname{Re}\!\big[ \operatorname{Tr}\!\big[ \hat{\varrho}^{[1]} \hat{a}_{k}\hat{a}_{l}\big]\big] - \operatorname{Tr}\!\big[ \hat{\varrho}^{[1]}\hat{a}_{k}\big] \operatorname{Tr}\big[ \hat{\varrho}^{[1]}\hat{a}_{l}\big]  \nonumber \\
&=&  \operatorname{Re}\!\big[ \langle\phi| \hat{a}_{k}\hat{a}_{l}|\phi\rangle\big] - \langle\phi|\hat{a}_{k}|\phi\rangle \langle\phi|\hat{a}_{l}|\phi\rangle.
\end{eqnarray}
Now, let $\mathbbm{1} = \sum_{n=1}^{D}|\xi_n\rangle\langle \xi_n|$ where $|\xi_1\rangle = |\phi\rangle$. With this
\begin{eqnarray}
\mathcal{I}^{[1]}_{k,l} &=&  \sum_{n=2}^{D} \operatorname{Re}\!\big[ \langle\phi| \hat{a}_{k}|\xi_{n}\rangle\langle\xi_{n}|\hat{a}_{l}|\phi\rangle\big] \nonumber\\
&&+ \operatorname{Re}\!\big[\langle\phi|\hat{a}_{k}|\phi\rangle \langle\phi|\hat{a}_{l}|\phi\rangle \big] - \langle\phi|\hat{a}_{k}|\phi\rangle \langle\phi|\hat{a}_{l}|\phi\rangle.\nonumber \\
&=& \sum_{n=2}^{D} \operatorname{Re}\!\big[ \langle\phi| \hat{a}_{k}|\xi_{n}\rangle\langle\xi_{n}|\hat{a}_{l}|\phi\rangle\big].
\end{eqnarray}
Next, we define vectors $\boldsymbol{x}_{n}\in\mathbb{C}^{D}$, $n=2,\ldots,D$, with entries $x_n^{k} = \langle\phi|\hat{a}_{k}|\xi_n\rangle$. With this, the QFIM $\mathcal{I} = 4N\mathcal{I}^{[1]}$ reduces to
\begin{equation*}
\mathcal{I} = 4N\sum_{n=2}^{D}\operatorname{Re}\!\big[ \boldsymbol{x}_{n}\boldsymbol{x}_{n}^{\dagger} \big] 
= 2N\sum_{n=2}^{D} \left[  \boldsymbol{x}_{n}\boldsymbol{x}_{n}^{\dagger} + \left( \boldsymbol{x}_{n}\boldsymbol{x}_{n}^{\dagger}\right)^{*}\right]
\end{equation*}
which is a sum of $2(D-1)$ rank one matrices. Hence, $\operatorname{rank}[\mathcal{I}] \leq 2(D-1)$.

\section{\label{app:UnitaryMultiParameterEstimationSLDs}Unitary multi-parameter estimation: \\Saturating the quantum Cram{\'e}r-Rao bound}
Next, we prove that the quantum Cram{\'e}r-Rao bound can be saturated in the setting we are considering. In a multi-parameter estimation setup, in general, the SLDs do not commute. This is the reason why the quantum Cram{\'e}r-Rao bound may not be saturated \cite{vidrighin2014,crowley2014}. As we will see, however, if the expectation value of the commutator vanishes, i.e., 
\begin{equation}
\langle\psi_{\boldsymbol{\varphi}}|\hat{L}_{k}\hat{L}_{l} - \hat{L}_{l}\hat{L}_{k}|\psi_{\boldsymbol{\varphi}}\rangle = 0,
\end{equation}
the bound can sill be achieved (see also \cite{Matsumoto2002}). One finds
\begin{equation}
\begin{split}
&\langle\psi_{\boldsymbol{\varphi}}|\hat{L}_{k}\hat{L}_{l}|\psi_{\boldsymbol{\varphi}}\rangle /4 \\
&= \langle\psi| \left( \hat{O}_{k}|\psi\rangle\langle\psi| + |\psi\rangle\langle\psi|\hat{O}_{k}^{\dagger} \right) \times \\
&\hspace{1.5cm}\times\left( \hat{O}_{l}|\psi\rangle\langle\psi| + |\psi\rangle\langle\psi|\hat{O}_{l}^{\dagger} \right)|\psi\rangle \\
&= \langle\psi|\hat{O}_{k}|\psi\rangle \langle\psi|\hat{O}_{l}|\psi\rangle + \langle\psi|\hat{O}_{k}|\psi\rangle \langle\psi|\hat{O}_{l}^{\dagger}|\psi\rangle \\
&\hspace{1.5cm}+\langle\psi|\hat{O}_{k}^{\dagger}\hat{O}_{l}|\psi\rangle + \langle\psi|\hat{O}_{k}^{\dagger}|\psi\rangle \langle\psi|\hat{O}_{l}^{\dagger}|\psi\rangle.
\end{split}
\end{equation}
With this, 
\begin{equation}
\begin{split}
&\langle\psi_{\boldsymbol{\varphi}}|\hat{L}_{k}\hat{L}_{l} - \hat{L}_{l}\hat{L}_{k}|\psi_{\boldsymbol{\varphi}}\rangle \\
&= 8\i\left[\operatorname{Im}[\langle\psi|\hat{O}_k|\psi\rangle  \langle\psi|\hat{O}_{l}^{\dagger}|\psi\rangle] + \operatorname{Im}[\langle\psi|\hat{O}_{k}^{\dagger}\hat{O}_{l}|\psi\rangle] \right] \\
&=8\i\operatorname{Im}\left[ \langle\psi|\hat{A}_{k}\hat{A}_{l}|\psi\rangle \right],
\end{split}
\end{equation}
where $\hat{O}_{k} = -\i\hat{A}_{k}$ and $\langle\psi|\hat{A}_{k}|\psi\rangle\in\mathbb{R}$ as $\hat{A}_{k} = \hat{A}_{k}^{\dagger}$. For $\hat{A}_{k}=\sum_{n=1}^{N}\hat{a}_{k}^{[n]}$ this expectation value  reduces to 
\begin{eqnarray}
&&\hspace{0.2cm} 8\i \!\!\sum_{n\neq m} \!\!\operatorname{Im}\!\!\left[\operatorname{Tr}\!\big[\hat{\varrho}^{[n,m]} \hat{a}_{k}^{[n]}\!\otimes\!\hat{a}_{l}^{[m]}\big]\!\right] \!\!+\! 8\i \!\sum_{n} \operatorname{Im}\!\!\left[\operatorname{Tr}\!\big[\hat{\varrho}^{[n]} \hat{a}_{k}^{[n]}\hat{a}_{l}^{[n]}\big]\!\right] \nonumber
\\ 
&&\hspace{0.2cm} =  8\i \!\sum_{n} \operatorname{Im}\!\!\left[\operatorname{Tr}\big[\hat{\varrho}^{[n]} \hat{a}_{k}^{[n]}\hat{a}_{l}^{[n]}\big]\right] \nonumber
\\ 
&&\hspace{0.2cm} = 8\i N \operatorname{Im}\!\!\left[ \operatorname{Tr}\big[ \hat{\varrho}^{[1]} \hat{a}_{k}\hat{a}_{l} \big] \right],
\label{eqn:ConditionForQuantumLimitLocal}
\end{eqnarray}
since $\operatorname{Tr}[\hat{\varrho}^{[n,m]}\hat{a}_{k}^{[n]}\otimes\hat{a}_{l}^{[m]}]\in \mathbb{R}$ for $n\neq m$ and the last equation is valid for permutational invariant systems.

Next, we prove that 
\begin{equation}
\operatorname{Im}\left[ \langle\psi|\hat{A}_{k}\hat{A}_{l}|\psi\rangle \right] = 0
\label{eqn:ConditionForCramerRaoBound}
\end{equation}
is a sufficient condition for the Cram{\'e}r-Rao bound to be saturated. First, note that each SLD $\hat{L}_{k}$ (see Eqn.~\eqref{eqn:SLDForSetting}) is of rank~$2$ where the non-zero eigenvalues are given by 
\begin{equation}
\lambda_k^{\pm} = \pm2\sqrt{\langle\psi|\hat{A}^{2}_{k}|\psi\rangle - \langle\psi|\hat{A}_{k}|\psi\rangle^2}
\end{equation}
with the corresponding eigenvectors
\begin{equation}
|\phi_k^{\pm}\rangle = a_k \hat{U}\hat{A}_{k}|\psi\rangle + b_{k}^{\pm} \hat{U}|\psi\rangle,
\end{equation}
where 
\begin{eqnarray*}
a_{k} &=&  \frac{1}{\sqrt{2\langle\psi|\hat{A}_{k}^2|\psi\rangle}}\;,\\
b^{\pm}_{k} &=& -\frac{\langle\psi|\hat{A}_{k}|\psi\rangle \pm \i \sqrt{\langle\psi|\hat{A}_{k}^2|\psi\rangle - \langle\psi|\hat{A}_{k}|\psi\rangle^2}}{\sqrt{2\langle\psi|\hat{A}_{k}^2|\psi\rangle}}.
\end{eqnarray*}
Hence, the eigenspaces of $\{\hat{L}_{k}\}$ are spanned by the $d+1$ vectors 
\begin{equation}
|\xi_{0}\rangle = \hat{U}|\psi\rangle,\;\;|\xi_{k}\rangle =\hat{U}\hat{A}_{k}|\psi\rangle \mbox{ for }k=1,\ldots,d.
\label{eqn:SetOfLinearlyIndependentVectors}
\end{equation}
Secondly, we show that these vectors are linearly independent, i.e., the subspace resulting by combining the eigenspaces of the SLDs is of dimension $d+1$. To prove this assertion, let $G\in\mathbb{R}^{(d+1)\times (d+1)}$ be the Gramian matrix of the vectors given in Eqn.~\eqref{eqn:SetOfLinearlyIndependentVectors}, i.e., $G_{k,l} = \langle\xi_{k}|\xi_{l}\rangle$. One finds
\begin{equation}
G \!=\! \begin{pmatrix} 
1 & \langle\psi|\hat{A}_{1}|\psi\rangle &\ldots& \langle\psi|\hat{A}_{d}|\psi\rangle \\
\langle\psi|\hat{A}_{1}|\psi\rangle & \langle\psi|\hat{A}_{1}\hat{A}_{1}|\psi\rangle  & \ldots & \langle\psi|\hat{A}_{1}\hat{A}_{d}|\psi\rangle \\
\vdots & & \ddots & \vdots\\
\langle\psi|\hat{A}_{d}|\psi\rangle & \langle\psi|\hat{A}_{d}\hat{A}_{1}|\psi\rangle &\ldots & \langle\psi|\hat{A}_{d}\hat{A}_{d}|\psi\rangle
\end{pmatrix},
\end{equation}
where, of course, the probe state $|\psi\rangle$ is normalised. It remains to show that the Gramian matrix has full rank. For this, recall that for every Hermitian matrix $M$ that can be partitioned as 
\begin{equation}
M = \begin{pmatrix} A & B \\ B^{\dagger} & C  \end{pmatrix},
\end{equation}
where $A$ and $C$ are square matrices, it holds that \cite{HornJ91}
\begin{equation}
M>0 \Leftrightarrow A>0 \mbox{ and } C-B^{\dagger}A^{-1}B >0,
\end{equation}
where $M>0$ denotes positive definiteness, i.e.,  $\langle x|M|x\rangle > 0$ for all $|x\rangle$. Note that $S = C-B^{\dagger}A^{-1}B$ is called the Schur complement of block $A$ of $M$. Now, let $A=1$, $B_{k} = \langle\psi|\hat{A}_{k}|\psi\rangle$, and $C_{k,l} = \langle\psi|\hat{A}_{k}\hat{A}_{l}|\psi\rangle$. Obviously, $A>0$. Further, the Schur complement is given by
\begin{equation}
S_{k,l} = \langle\psi|\hat{A}_{k}\hat{A}_{l}|\psi\rangle - \langle\psi|\hat{A}_{k}|\psi\rangle \langle\psi|\hat{A}_{l}|\psi\rangle, 
\end{equation}
i.e., $S = \mathcal{I}(\boldsymbol{\varphi}) / 4$ given that the expectation values of all commutators of the SLDs vanish, i.e., $\operatorname{Im}[ \langle\psi|\hat{A}_{k}\hat{A}_{l}|\psi\rangle] = 0$, see Eqns.~\eqref{eqn:QFIwithAoperator_Appendix} and \eqref{eqn:ConditionForCramerRaoBound}. As we assume that the QFIM has full rank (with positive eigenvalues), we have $S>0$.  Thus, the Gramian matrix is positive definite and, hence, has full rank such that the set of vectors given in Eqn.~\eqref{eqn:SetOfLinearlyIndependentVectors} is linearly independent. Hence, one can find an orthogonal basis of the subspace spanned by the eigenvectors of all SLDs by a Gram-Schmidt orthogonalisation procedure starting with the vector $|\xi_{0}\rangle = \hat{U}|\psi\rangle$. The $d+1$ projectors onto these orthogonal vectors, together with one element that accounts for the normalisation, form a set of POVMs of cardinality $d+2$. As one element of this POVM is the projector onto the time-evolved probe state, the results of Ref. \cite{humphreys2013} prove that this set of POVMs saturates the quantum Cram{\'e}r-Rao bound.

\section{\label{app:dg3}$d > 3$}
Let us restrict to two-level systems and come back to the setting where the task is to estimate the three components of a magnetic field pointing in an arbitrary direction, i.e., the evolution under the Hamiltonian
\begin{equation}
\hat{h} = \boldsymbol{\hat{\mu}} \boldsymbol{B} = \sum_{k=1}^{3} \hat{\mu}_k B_{k} = \sum_{k=1}^{3} \frac{\mu}{2}B_{k}\hat{\sigma}_{k} := \sum_{k=1}^{3}\varphi_k \hat{\sigma}_{k}.
\end{equation}
It is worth mentioning that $d=3$ is the maximal number of to-be-estimated parameters given the Hamiltonian acts on each site independently. Assume that 
\begin{equation}
\hat{h} = \sum_{k=1}^{d} \varphi_k \hat{h}_{k}
\end{equation}
for a $d>3$. We can always decompose each $\hat{h}_{k}$ in the (normalised) Pauli basis $\{\hat{P}_{i}\}$ with $\hat{P}_{1} = \hat{\sigma}_1/\sqrt{2}$, $\hat{P}_{2} = \hat{\sigma}_2/\sqrt{2}$, $\hat{P}_{3} = \hat{\sigma}_3/\sqrt{2}$, and $\hat{P}_{4} =\mathbbm{1}/\sqrt{2}$. One finds
\begin{equation}
\hat{h} = \sum_{l=1}^{4} \left( \sum_{k=1}^{d} \varphi_k \operatorname{tr}[\hat{P}_{l}\hat{h}_{k}]\right)\hat{P}_{l} = \sum_{l=1}^{4} c_l \hat{P}_{l}
\end{equation}
such that, in fact, $\{c_l\}$ are the independent parameters (and the parameters $\{\varphi_k\}$ are determined by the $\{c_k\}$). Further, any contribution that is proportional to the identity can be neglected as this would result in an unobservable global phase. Hence, estimating these three phases can be interpreted as single-particle Hamiltonian tomography at the Heisenberg limit.

\section{\label{app:SingleParameterEstimation}Single-parameter estimation and multi-parameter estimation with commuting generators}

Let us first review the results for single-parameter estimation~\cite{Giovannetti2006} in the framework discussed in the main text. For this, let the single particle Hamiltonian governing the time evolution be given by $\hat{h} = \varphi \hat{h}_{1} + \hat{h}_{2}$ where the Hermitian operators $\hat{h}_{1}$ and $\hat{h}_{2}$ do not necessarily commute. Note that this includes the estimation of one direction of a magnetic field pointing in an arbitrary direction where the remaining directions are kept constant, e.g., $\hat{h} = \varphi_{x}\hat{\sigma}_{x} + \hat{h}_{2}$ with $\operatorname{tr}[\hat{\sigma}_{x} \hat{h}_{2}] = 0$ and $[\hat{\sigma}_{x},\hat{h}_{2}]\neq 0$. As we allow to probe the magnetic field with $N$ particles simultaneously, the unitary evolution is given by
\begin{equation}
\hat{U} = \bigotimes_{n=1}^{N}  \e^{-\i \hat{h}^{[n]}} = \prod_{n=1}^{N} \e^{-\i (\varphi \hat{h}^{[n]}_{1} + \hat{h}_{2}^{[n]})} = \e^{-\i \varphi \hat{H}}
\end{equation}
with $\hat{H}=\sum_{n=1}^{N} (\varphi\hat{h}_{1}^{[n]}+\hat{h}_{2})$ the $N$-particle Hamiltonian. Hence, $\hat{A} = \sum_{n=1}^{N}\hat{a}^{[n]}$ where
\begin{equation}
\hat{a}^{[n]} = \int_{0}^{1} d\alpha\;\e^{\i \alpha (\varphi \hat{h}^{[n]}_{1} + \hat{h}_{2}^{[n]})} \hat{h}_{1}^{[n]}\e^{-\i \alpha (\varphi \hat{h}^{[n]}_{1} + \hat{h}_{2}^{[n]})},
\end{equation}
such that 
\begin{equation}
\mathcal{I}(\varphi) = 4(\langle \psi| \hat{A}^2|\psi\rangle - \langle\psi| \hat{A}|\psi\rangle^2). 
\end{equation}
Now, for product probe states of the form $|\psi\rangle = \bigotimes_{n=1}^{N}|\phi_n\rangle$, one finds 
\begin{equation}
\mathcal{I} = 4 \sum_{n=1}^{N} \left(\langle\phi_n|\big(\hat{a}^{[n]}\big)^2|\phi_n\rangle - \langle\phi_n|\hat{a}^{[n]}|\phi_n\rangle^2\right) 
\end{equation}
which reduces to 
\begin{equation}
\mathcal{I} = 4N \left(\langle\phi|\hat{a}^2|\phi\rangle - \langle\phi|\hat{a}|\phi\rangle^2\right) 
\end{equation}
given that $|\phi_n\rangle = |\phi\rangle$ for all $n=1,\ldots,N$. The latter is maximised by states of the form $|\phi\rangle = (|\phi^{\text{max}}\rangle + |\phi^{\text{min}}\rangle)/\sqrt{2}$ where $\{|\phi^{\text{min}}\rangle,|\phi^{\text{max}}\rangle\}$ are the eigenstates of $\hat{a}$ corresponding to the minimal $\lambda_{\text{min}}(\hat{a})$ and maximal $\lambda_{\text{max}}(\hat{a})$ eigenvalue. With this,  
\begin{equation}
\mathcal{I} = N (\lambda_{\text{max}}(\hat{a}) - \lambda_{\text{min}}(\hat{a}))^2. 
\end{equation}
Allowing for entangled probe states $|\psi\rangle$, it is well known that the maximal quantum Fisher information is obtained by using GHZ-type states \cite{Giovannetti2006}, i.e., 
\begin{equation}
|\Phi_{\hat{a}}\rangle = [|\phi^{\text{max}}\rangle^{\otimes N} + |\phi^{\text{min}}\rangle^{\otimes N}]/\sqrt{2}.
\label{eqn:GHZstate}
\end{equation}
Note that $\{|\phi^{\text{max}}\rangle^{\otimes N},|\phi^{\text{min}}\rangle^{\otimes N}\}$ are the eigenstates of $\hat{A}$ corresponding to its maximal and minimal eigenvalue, i.e., $\{N\lambda_{\text{max}}(\hat{a}),N\lambda_{\text{min}}(\hat{a})\}$.
With this, 
\begin{equation}
\mathcal{I} = N^2(\lambda_{\text{max}}(\hat{a})-\lambda_{\text{min}}(\hat{a}))^2.
\end{equation}
Moreover, the Cram{\'e}r-Rao bound can always be attained yielding the quantum advantage of a Heisenberg scaling in contrast to the shot noise limit with respect to the precision of the parameter $\varphi$. Note that for $\hat{h}_{2}=0$ and $\hat{h}_{1} = \hat{\sigma}_{k}$, with either $k=1,2$, or $3$, this reduces to the scenario of estimating the magnetic field when the direction (here X, Y, or Z) is known.

Next, let us discuss a setting for multi-parameter estimation where the generators  $\{\hat{H}_{k}\}$ of the unitary time evolution commute, i.e., where 
\begin{equation}
\hat{U} = \e^{-\i \hat{H}} \text{ with } \hat{H}(\boldsymbol{\varphi}) = \sum_{k=1}^{d}\varphi_k \hat{H}_{k}
\end{equation}
and $[\hat{H}_{k},\hat{H}_{l}]=0$ for all $k,l=1,\ldots,d$. For this, we review the results obtained in \cite{humphreys2013} in the framework discussed in the main text. Recall that in \cite{humphreys2013} the task is to estimate $d$ phases in a $d+1$-mode interferometer. Each phase is independently imprinted on the probe state in one mode of the interferometer, whereas the remaining mode serves as a reference. This is done via the generators $\hat{H}_{k} = \hat{N}_{k}$ where $\hat{N}_{k}$ is the number operator for mode $k$. With this 
\begin{equation}
\hat{U}(\boldsymbol{\varphi}) = \e^{-\i \sum_{k=1}^{d} \varphi_k \hat{N}_{k}}.
\end{equation}
Further, as $[\hat{N}_{k},\hat{N}_{l}]=0$, one finds $\hat{A}_{k}(\boldsymbol{\varphi}) = \hat{N}_{k}$ such that the quantum Fisher information matrix is given by 
\begin{equation}
\mathcal{I}_{k,l} = 4  \operatorname{Re}\!\left[ \langle \psi|  \hat{N}_k\hat{N}_l|\psi \rangle -  \langle\psi|\hat{N}_k|\psi\rangle  \langle\psi|\hat{N}_l|\psi\rangle \right]. 
\end{equation}
The probe state for this QFIM presented in \cite{humphreys2013} results from the same intuition as the probe state discussed in the main text for the magnetic field estimation: $|\psi\rangle$ is a superposition of the states that yield a quantum advantage when estimating the parameters individually. While we cannot present a proof that this intuition is optimal, it seems a good first guess when considering simultaneous multi-parameter estimation.

Finally, as the generators $\{\hat{N}_k\}$ commute, one finds $\langle \psi|\hat{N}_{k}\hat{N}_{l}|\psi\rangle^{*} = \langle \psi|\hat{N}_{k}\hat{N}_{l}|\psi\rangle$ such that $\operatorname{Im}\big[ \langle\psi|\hat{N}_{k}\hat{N}_{l}|\psi\rangle \big] = 0$ and the Cram{\'e}r-Rao bound can be saturated.

\section{\label{app:Reduceddensitymatricesoftheprobestate}Reduced density matrices of the probe state}
In this section of the appendix, we discuss the reduced density matrices of the probe state given by 
\begin{equation}
\begin{split}
|\psi\rangle  &=\mathcal{N}\!\left( |\Phi_1\rangle + |\Phi_2\rangle + |\Phi_3\rangle \!\right) \\
&=\mathcal{M}\,\sqrt{2}\,\left( |\Phi_1\rangle + |\Phi_2\rangle + |\Phi_3\rangle \!\right) \\
&= \mathcal{M}\left(\right. |\phi^{+}_{1}\rangle^{\otimes N} \!+\!|\phi^{-}_{1}\rangle^{\otimes N} \\
 &\hspace{0.7cm}+|\phi^{+}_{2}\rangle^{\otimes N} \!+\! |\phi^{-}_{2}\rangle^{\otimes N} \!+\! |\phi^{+}_{3}\rangle^{\otimes N}  \!+\! |\phi^{-}_{3}\rangle^{\otimes N}\!\left.\right), \\
\end{split}
\label{eqn:ProbeStateAppendix}
\end{equation}
where $|\phi^{\pm}_{k}\rangle$ is the eigenvector of the Pauli operator $\hat{\sigma}_{k}$ corresponding to the eigenvalue $\pm1$ for all $k=1,2,3$ and we defined $|\Phi_k\rangle = \bigl(|\phi_{k}^{+}\rangle^{\otimes N} + |\phi_{k}^{-}\rangle^{\otimes N}\bigr)/\sqrt{2}$. First, note that the normalisation constant is determined via
\begin{eqnarray}
&1&=\mathcal{M}^2\left[ 6 + 4\left(\frac{1+\i}{2}\right)^N +  4\left(\frac{1-\i}{2}\right)^N  \right. \\
&&+ \left.10\left(\frac{1}{\sqrt{2}}\right)^N + 2\left(\frac{-1}{\sqrt{2}}\right)^N + 2\left(\frac{\i}{\sqrt{2}}\right)^N + 2\left(\frac{-\i}{\sqrt{2}}\right)^N \right]. \nonumber 
\end{eqnarray}
Hence, $\mathcal{M}\rightarrow 1/\sqrt{6}$ for $N\rightarrow \infty$. Next, let us analyse the two-body reduced density matrix. First, note that 
\begin{equation}
\begin{split}
\hat{\varrho}^{[2]}_{k} &= \operatorname{tr}_{\backslash 2}\left[ |\Phi_k\rangle\langle\Phi_k| \right] \\
&= \frac{1}{2}\left( |\phi^{+}_{k},\phi^{+}_{k}\rangle\langle \phi^{+}_{k},\phi^{+}_{k}| \!+\! |\phi^{-}_{k},\phi^{-}_{k}\rangle\langle \phi^{-}_{k},\phi^{-}_{k}| \right) \\ 
&= \frac{1}{4}\left( \mathbbm{1}_2\otimes\mathbbm{1}_2 + \hat{\sigma}_{k}\otimes \hat{\sigma}_{k}\right)
\end{split}
\end{equation}
for all $k$. Moreover, terms like $\operatorname{tr}_{\backslash 2}\left[ |\Phi_k\rangle\langle\Phi_l| \right]$ scale as $1/2^{N/2}$ such that for $N\rightarrow \infty$ they vanish. Hence, in the limit $N\rightarrow \infty$, the two-body marginal of the probe state converges to 
\begin{equation}
\hat{\varrho}^{[2]} = \operatorname{tr}_{\backslash 2}\left[ |\phi\rangle\langle\phi| \right] \rightarrow \frac{1}{3}\sum_{k=1}^{3} \hat{\varrho}_{k}^{[2]}.
\end{equation}
Finally, let us note that for $N=8n$, $n\in\mathbb{N}$, this is exact, i.e., $\hat{\varrho}^{[2]} \equiv \sum_{k=1}^{3} \hat{\varrho}_{k}^{[2]}/3$.

\section{\label{app:DerivationQFI}Derivation of the QFIM}
Here, we calculate the QFIM for the probe state given in Eqn.~\eqref{eqn:ProbeStateAppendix} with one- and two-body reduced density matrices $\hat{\varrho}^{[1]} = \mathbbm{1}_{2}/2$ and $\hat{\varrho}^{[2]} = \mathbbm{1}_{2}\otimes\mathbbm{1}_{2}/4 + \sum_{k=1}^{3}\hat{\sigma}_{k}\otimes\hat{\sigma}_{k}/12$, respectively. We begin by noting that $\operatorname{Tr}[\hat{a}_{k}] = 0, \forall k,$ since Pauli operators are traceless and, hence, $\mathcal{I}_{k,l}^{[1]}(\boldsymbol{\varphi}) = \operatorname{Tr}\left[ \hat{a}_{k}\hat{a}_{l} \right]/2$. With this
\begin{eqnarray}
\mathcal{I}^{[2]}_{k,l}(\boldsymbol{\varphi}) &=& \operatorname{Tr}\big[ \hat{\varrho}^{[2]} \hat{a}_{k}\otimes\hat{a}_{l}\big] - \operatorname{Tr}\big[ \hat{\varrho}^{[1]}\hat{a}_{l}\big] \operatorname{Tr}\big[ \hat{\varrho}^{[1]}\hat{a}_{l}\big] \nonumber \\
&=& \frac{1}{12} \sum_{m=1}^{3} \operatorname{Tr}\left[ (\hat{\sigma}_m\otimes \hat{\sigma}_m) (\hat{a}_{k}\otimes\hat{a}_{l})\right]  \nonumber \\
&=& \frac{1}{6} \operatorname{Tr}\left[\sum_{m=1}^{3} \operatorname{Tr}\left[\hat{P}_{m}\hat{a}_{k}\right]\hat{P}_{m}\hat{a}_{l}\right] =\frac{1}{6} \operatorname{Tr}\left[ \hat{a}_{k}\hat{a}_{l} \right] \nonumber
\end{eqnarray}
as $\hat{P}_{k}=\hat{\sigma}_k/\sqrt{2}$, $k=1,2,3$, together with $\hat{P}_{4}=\mathbbm{1}/\sqrt{2}$ is an orthonormal basis and the contribution proportional to $\hat{P}_{4}$ for the operator $\hat{a}_{k}$ is zero. 
 Thus, 
\begin{equation}
\mathcal{I}^{[2]}_{k,l}(\boldsymbol{\varphi}) \!=\! \frac{1}{3} \operatorname{Tr}\left[\hat{\varrho}^{[1]} \hat{a}_{k}\hat{a}_{l} \right] \!=\! \frac{1}{3}\;\mathcal{I}_{k,l}^{[1]}(\boldsymbol{\varphi}).
\end{equation}
Hence, the QFIM is
\begin{equation}
\mathcal{I}_{k,l}(\boldsymbol{\varphi}) \!=\! \frac{4N(N+2)}{3} \; \mathcal{I}_{k,l}^{[1]} = \frac{2N(N+2)}{3} \operatorname{Tr}\left[\hat{a}_{k}\hat{a}_{l} \right].
\end{equation}

Using the definition of the operators $\{\hat{a}_{k}\}$, see Eqn.~\eqref{eqn:AOperatorLocal}, we have
\begin{equation}
\begin{split}
\operatorname{Tr}\left[ \hat{a}_{k}\hat{a}_{l} \right] &=  \!\int_0^1\!\!\! d\alpha\,d\beta\,\operatorname{Tr}\left[\e^{\i\alpha\hat{h}} \hat{\sigma}_{k} \e^{-\i\alpha\hat{h}}\e^{\i\beta\hat{h}}\hat{\sigma}_{l}\e^{-\i\beta\hat{h}}\right] \\
&=\!\int_0^1 \!\!\! d\alpha\,d\beta\,\operatorname{Tr}\left[ \hat{\sigma}_{l}\e^{\i(\alpha-\beta)\hat{h}}\hat{\sigma}_{k}\e^{-\i(\alpha-\beta)\hat{h}} \right] \\
&=\operatorname{Tr}\left[ \hat{\sigma}_l \hat{W}_{k} \right]
\end{split}
\end{equation}
such that the entries of $\mathcal{I}(\boldsymbol{\varphi})$ are given in terms of the entries of the operators 
\begin{equation}
\hat{W}_{k} = \!\int_0^1 \!\!\! d\alpha\,d\beta\,\e^{\i(\alpha-\beta)\hat{h}}\hat{\sigma}_{k}\e^{-\i(\alpha-\beta)\hat{h}}
\label{eqn:WOperators}
\end{equation}
in the Pauli basis. To find analytic expression of these operators, recall that with $\|\boldsymbol{n}\|^2 = 1$ one has
\begin{equation}
\e^{-\i\theta (\sum_{k=1}^{3}n_k \hat{\sigma}_k)} = \cos[\theta]\mathbbm{1} - \i\sin[\theta]\sum_{k=1}^{3}n_k\hat{\sigma}_k.
\end{equation}
Now, let
\begin{equation}
\xi = \sqrt{\varphi_1^2 + \varphi_2^2 + \varphi_3^2} \;\text{ and }\; \eta_k = \frac{\varphi_k}{\sqrt{\varphi_1^2 + \varphi_2^2 + \varphi_3^2}}
\end{equation}
for all $k=1,2,3$ (corresponding to $X,Y$ und $Z$). We find for the operators $\hat{W}_{k}$
\begin{equation*}
\begin{split}
\hat{W}_{1} = &\hat{\sigma}_1 \left[ 1 + \sinc^2[\xi] +(1-\sinc^2[\xi])(\eta_1^2-\eta_2^2-\eta_3^2)\right] / 2 \\
+&\hat{\sigma}_2\left[ 1-\sinc^2[\xi] \right]\eta_1\eta_2\\
+&\hat{\sigma}_3\left[ 1-\sinc^2[\xi]\right]\eta_1\eta_3,
\end{split}
\end{equation*}
where $\sinc[\xi] = \sin[\xi]/\xi$. Further,
\begin{equation*}
\begin{split}
\hat{W}_{2} = &\hat{\sigma}_1\left[ 1-\sinc^2[\xi] \right]\eta_1\eta_2\\
+&\hat{\sigma}_2 \left[ 1 + \sinc^2[\xi] +(1-\sinc^2[\xi])(-\eta_1^2+\eta_2^2-\eta_3^2)\right] / 2 \\
+&\hat{\sigma}_3\left[ 1-\sinc^2[\xi]\right]\eta_2\eta_3,
\end{split}
\end{equation*}
and
\begin{equation*}
\begin{split}
\hat{W}_{3} = &\hat{\sigma}_1\left[ 1-\sinc^2[\xi] \right]\eta_1\eta_3\\
+&\hat{\sigma}_2\left[ 1-\sinc^2[\xi]\right]\eta_2\eta_3 \\
+&\hat{\sigma}_3 \left[ 1 + \sinc^2[\xi] +(1-\sinc^2[\xi])(-\eta_1^2-\eta_2^2+\eta_3^2)\right] / 2. \\
\end{split}
\end{equation*}
With this, the QFIM simplifies to
\begin{equation}
\mathcal{I}_{k,l} \!=\! \frac{4}{3}N(N+2)\!\left[ (1\!-\!\sinc^2[\xi]) \eta_k\eta_l + \delta_{k,l} \sinc^2[\xi]\right].
\end{equation}


\begin{thebibliography}{00}
\bibitem{remark1}
Note that a more general argument can be constructed when $\hat{\varrho}^{[1]} = \sum_i p_i |\psi_i\rangle\langle\psi_i|$ where $\{|\psi_{i}\rangle\}$ are the eigenvectors of $\hat{a}_{k}$ or $\hat{a}_{l}$, respectively.
\bibitem{remark2}
Note that for $\hat{\varrho}^{[1]}=\mathbbm{1}_2/2$ the adjustable local phases $\delta_k$, $k=1,2,3$, do not depend on $\varphi_k$, nor on the number of sub-systems $N$. For even $N$, i.e., $N=2n$, $n\in\mathbb{N}$, one possible choice is $\delta_1=0$, $\delta_2=\pi/2$, and $\delta_3=0$. Moreover, for $N=4n$, $n\in\mathbb{N}$, another possible realisation is given by $\delta_1=\delta_2=\delta_3=0$.
\bibitem{remark3}
For $N=2n$, $n\in\mathbb{N}$, and $\delta_1=0$, $\delta_2=\pi$, and $\delta_3=\pi$ the states $|\Psi_k\rangle$ are orthogonal and hence $\hat{\Pi}_{k}^{(1)}$ is a valid set of POVMs. Moreover, for $N=4n$, $n\in\mathbb{N}$, another possible choice is $\delta_1=\delta_2=\delta_3=\pi$ (used for the numerics).
\end{thebibliography}
\end{document}